\documentclass[,twocolumn,,trackchanges]{aastex631}
\usepackage{amssymb}
\usepackage{amsmath}
\usepackage{epstopdf}
\usepackage{natbib}
\usepackage{xcolor}
\usepackage{url}
\usepackage{epsfig}
\usepackage[mathscr]{euscript}
\usepackage{threeparttable}
\begin{document}

\linespread{1.12}\selectfont
\title{The Multiband Emission of the two-component Gamma-Ray Burst jet influenced by progenitor winds within the Accretion Disk of Active Galactic Nuclei}

\correspondingauthor{Wei-Hua Lei}
\email{leiwh@hust.edu.cn}

\author{Hao-Yu Yuan}
\affiliation{Department of Astronomy, School of Physics, Huazhong University of Science and Technology, Luoyu Road 1037, Wuhan, 430074, China}

\author[0000-0003-3440-1526]{Wei-Hua Lei}
\affiliation{Department of Astronomy, School of Physics, Huazhong University of Science and Technology, Luoyu Road 1037, Wuhan, 430074, China}

\begin{abstract}

Gamma-ray bursts (GRBs), both from  merger of binary compact objects (short GRBs) and collapse of massive stars (long GRBs), are expected to occur in the dense environments, e.g., the accretion disk of active galactic nuclei (AGN). The propagating of GRB jets in such dense environment will result in multiband transients. Investigating the properties of these transients plays important roles in their identification, understanding the jet structure and constraining population of the star and compact object in AGN disks. In this work, we intend to study the propagation and emission of a two-component GRB jet (a fast narrow component and a wide slow one) in the AGN disk. We consider the influence of wind from the short and long GRB progenitors, which would reconstruct the surrounding density distribution and form a cavity in the AGN disk.
We find that the long GRB jets will be choked, the dynamcis and the emission are resemble to the case without cavity. The cocoon breakout emission can be detected by EP and HXMT. For short GRBs, we expect a non-thermal afterglow emission from the narrow and wide jet (if it is more energetic than the narrow one) and a cocoon breakout emission from the choked wide jet, which can be monitored by EP and HXMT, respectively. Therefore, the joint observations by EP and HXMT might be helpful to distinguish the type of GRBs in the AGN disk and the jet components.
\end{abstract}

\keywords{\href{http://astrothesaurus.org/uat/629}{gamma ray burst (629)}, \href{http://astrothesaurus.org/uat/562}{Galaxy accretion disks (562)}}


\section{Introduction}\label{Intro}
The activity of galactic nuclei (AGN) is believed to be fueled by the accretion of dense matter of the disk onto the central supermassive black hole (SMBH) \citep{Lynden-Bell(1969),ShakuraSunyaev(1973),SirkoGoodman(2003)}.
The recent observations suggest that AGN disks constitute dense environments harboring stars and compact objects.
Recently, the Laser Interferometer Gravitational Wave Observatory (LIGO) has successfully detected a gravitational wave (GW) signal GW190521 from the merger of binary black holes (BBHs) \citep{Abbottetal.(2020)}. Typically, BBH mergers do not produce electromagnetic (EM) counterparts unless there is an accompanying presence of gas \citep{McKernanetal.(2019)}.
However, approximately 34 days following the GW190521 trigger, the Zwicky Transient Facility (ZTF) detected a potential optical electromagnetic counterpart, ZTF19abanrhr, which exhibited consistency with a plausible association between the GW event and the AGN J14942.3+344929
\citep{Grahametal.(2020),Ashtonetal.(2021), NitzCapano(2020)} \footnote{Recently, the association between GW190521 and ZTF19abanrhr was reanalyzed by \citet{Mortonetal.(2023)} within the Bayesian framework, utilizing the GWTC-2.1 data release \citep{Abbottetal.(2024)}, and found that the two events may be preferred over a random coincidence.}. More recently, \citet{Grahametal.(2023)} searched for possible EM counterparts to BBH mergers detected by LIGO/Virgo in O3 with ZTF. Based on ZTF sky coverage, AGN geometry, and merger
geometry, they expect that $\sim 3$ EM counterparts associated with GW events in O3.
In addition, through multi-frequency analysis, \citet{Levanetal.(2023)} suggested that GRB 191019A is likely a burst from dynamical interactions via compact object mergers in the AGN disk.

Stars can be developed inside AGN disks in two ways: gravitational instability within the disk \citep{Paczynski(1978), Goodman(2003), DittmannMiller(2020)}, or capture from the surrounding star cluster \citep{Artymowiczetal.(1993), Fabjetal.(2020)}. The death of massive stars would produce compact objects, such as neutron stars (NSs) and black holes (BHs). Gamma-ray burst (GRB) jets can be generated in the accretion disk due to the collapse of massive stars or the merger of binary compact objects \citep[BCOs]{BranchWheeler(2017), Woosley(1993)}. The connections of long duration GRBs (LGRBs) with core-collapse of massive stars, and short duration GRBs (SGRBs) with the mergers of BCOs have been supported by observations \citep[for a review]{WoosleyBloom(2006),Nakar2007,Zhang2018}.
\citet{Zhuetal.(2021)} have analyzed the jet produced by the merging of binary neutron stars in the AGN disk. They pointed out that the jet can not successfully break through the surface of the disk and will develop a hot cocoon, which can produce X-ray radiation brighter than the disk itself when the cocoon breaks through the disk surface.
\citet{Pernaetal.(2021)} argued that the GRB jet is harder to break through the disk around a more massive central SMBH. In addition to cocoon shock breakout, shock emission due to the interaction between jet and disk material will diffuse in the disk and produce supernova-like radiation as it emerges from the dense gas \citep{Chatzopoulosetal.(2012), Lietal.(2023)}.
The detailed inspection of these transients would be beneficial in identifying their physical origin and calibrating the populations of stars and compact objects in the AGN disk environment, which is of special relevance for the AGN channel of the LIGO/Virgo detections \citep{McKernan2020,Wang2021}, as well as for the stellar evolution in the AGN environments \citep{Cantiello2021,Jermyn2021,Dittmann2021}.

The studies mentioned above exclusively employ a simplistic uniform jet model \citep{Zhuetal.(2021), Pernaetal.(2021)}.
However, theoretical and observational studies revealed that at least some GRB jets could be structured \citep{Zhangetal.(2004), Meszarosetal.(1998), DaiGou(2001)}. Among which, the two-component jet has been widely mentioned \citep{Bergeretal.(2003), Shethetal.(2003), Huang2004, Wu2005, Xieetal.(2012), Xuetal.(2000),Yuanetal.(2024)}, e.g., the "brightest of all time" burst GRB 221009A \citep{Zhengetal.(2023)} and GRB 240529A \citep{Sunetal.(2024)}, which contain a narrow, highly relativistic component and a wider, moderately relativistic component.
Exploring the emergent emission of GRBs with two-component jets within AGN disks holds promise for enhancing our comprehension of nuclear transients.

On the other hand, both GRB progenitors (massive star and BCOs) will produce a wind. This wind interacts with the matter in the disk, exerting pressure and causing it to reach an equilibrium at some distance (form a low-density cavity). The influence of the winds on the dynamics and emission of GRB in AGN disk should be seriously considered. \citet{Lietal.(2023)} analyzed the effect of the wind cavity, but for core-collapse supernovae explosion. For SGRBs, the super-Eddington accretion of the compact objects in a dense environment might lead to a strong wind \citep{Jiangetal.(2014), Hashizumeetal.(2015)}, which may penetrate the AGN disk. \citet{YuanChenChaoetal.(2022)} argued that a successful GRB jet is expected in such a low-density cavity, producing significant GeV emission. Considering the difference in progenitor wind of LGRB and SGRB, the detailed investigation of the wind effects might provide possible clues to identify the type of GRBs from these transients.

Inspired by these considerations, in this paper, we intend to study the properties of multiband emission from a GRB with two-component jet in AGN disk environment by considering the effects of progenitor wind. The paper is organized as follows. In Section \ref{Cavity}, we describe the cavity induced by the progenitor wind of SGRB and LGRB, respectively. In Section \ref{disk GRB}, we present the dynamics and radiation of a GRB with a two-component jet launched in the AGN disk cavity. The results are presented in Section \ref{results}. Finally, Section \ref{Conclusions} summarizes and discusses our results.

\section{GRB progenitor wind and cavity induced in AGN disk} \label{Cavity}
The environment for GRBs in AGN disk differs significantly from interstellar medium (ISM). The dense gas suppresses the GRB jet, leaving a cocoon in the disk \citep{Zhuetal.(2021)}. On the other hand, the GRB progenitor wind interacts with the disk material, reconstructing the surrounding density configuration. In this section, we will investigate the surrounding environment for SGRBs and LGRBs in AGN disk, respectively.

\subsection{The AGN Disk Structure}\label{Sec：AGN disk}
We adopt the standard AGN disk model described in \citet{SirkoGoodman(2003)}, which includes the gravitational instability and auxiliary heating from nuclear fusion in the disk stars. The $\alpha$ prescription for the viscosity is adopted. The total luminosity of AGN disk radiation is related to the accretion rate $\dot{M}_{\rm SMBH}$ as $L_0=\epsilon_{\rm d} \dot{M}_{\rm SMBH}c^2$, and $l_{\rm E}=L_0/L_{\rm Edd}$ is the ratio of the accretion luminosity and the Eddington luminosity. the Eddington luminosity $L_{\rm Edd}=4\pi GM_{\rm SMBH}m_{\rm p}c/\sigma_{\rm T}$, where $G$ is the gravitational constant, $M_{\rm SMBH}$ is the mass of SMBH, $m_{\rm p}$ is the mass of proton and $\sigma_{\rm T}$ is thomson cross section.
The inner ($R_{\rm min}$) and outer($R_{\rm max})$ boundaries of the accretion disk are set to $R_{\rm min}=R_{\rm s}/(4\epsilon_{\rm d})$ and $R_{\rm max}=10^6 R_{\rm s}$, respectively, where $R_{\rm s}=\frac{2GM_{\rm SMBH}}{c^2}$ is the Schwarzschild radius. In the following calculation, we use $\alpha=0.01$, $\epsilon_{\rm d}=0.1$, $M_{\rm SMBH}=10^8M_\odot$ and $l_{\rm E}=0.5$.

The inner disk is powered entirely by gravitational energy, and the relevant equations of the structure are given by \citep{SirkoGoodman(2003)},
\begin{equation}
\label{eq:T_d_eff}
\sigma_{\rm SB}T_{\rm d,eff}^4=\frac{3}{8\pi}\dot{M}'\Omega_{\rm d}^2,
\end{equation}

\begin{equation}
\label{eq:T_d}
T_{\rm d}^4=\left(\frac{3}{8}\tau_{\rm d}+\frac{1}{2}+\frac{1}{4\tau_{\rm d}}\right)T_{\rm d,eff}^4,
\end{equation}

\begin{equation}
\label{eq:c_s_d}
c_{\rm s}^2\Sigma_{\rm d}=\frac{\dot{M}'\Omega_{\rm d}}{3\pi\alpha},
\end{equation}

\begin{equation}
\label{eq:P_d_rad}
c_{\rm s}^2=\frac{P_{\rm gas}+P_{\rm rad}}{\rho_{\rm d}},
\end{equation}
where $T_{\rm d,eff}$ is the effective temperature , $T_{\rm d}$ is the mid-disk temperature, $\tau_{\rm d}=\frac{\kappa_{\rm d}\Sigma_d}{2}$ is the optical depth from the middle to the surface of AGN disk, $\kappa_{\rm d}$ is the opacity of the middle disk, $\rho_{\rm d}$ is the mid-plane density, $\Sigma_{\rm d}=2\rho_{\rm d} H$ is the surface density, $H=\frac{c_{\rm s}}{\Omega_{\rm d}}$ is the scale height of the disk, $P_{\rm rad}=\frac{\tau_{\rm d}\sigma_{\rm SB}}{2c}T_{\rm d,eff}^4$ is the mid-disk radiation pressure, $P_{\rm gas}=\frac{\rho_{\rm d} k_B T_{\rm d}}{m_{\rm d}}$ is the mid-disk gas pressure, $\Omega_{\rm d}=(GM_{\rm SMBH}/R^3)^{1/2}$ is the angular frequency, $\dot{M}'=\dot{M}_{\rm SMBH}(1-\sqrt{R_{\rm min}/R})$, $m_{\rm d}=0.62m_{\rm p}$ is the mean molecular mass and $k_{\rm B}$ is the Boltzmann constant. We assume the viscosity is proportional to total pressure within above equations.
In order to solve these equations, we need to know the opacity ($\kappa_{\rm d}=\kappa_{\rm d}(\rho_{\rm d},T_{\rm d}$)) of the disk. We also use the opacity tables of \citet{IglesiasRogers(1996)} for high temperatures and \citet{AlexanderFerguson(1994)} for low temperatures, with $X = 0.70, Z = 0.03$.

\begin{figure*}
\centering
\includegraphics [angle=0,scale=0.2] {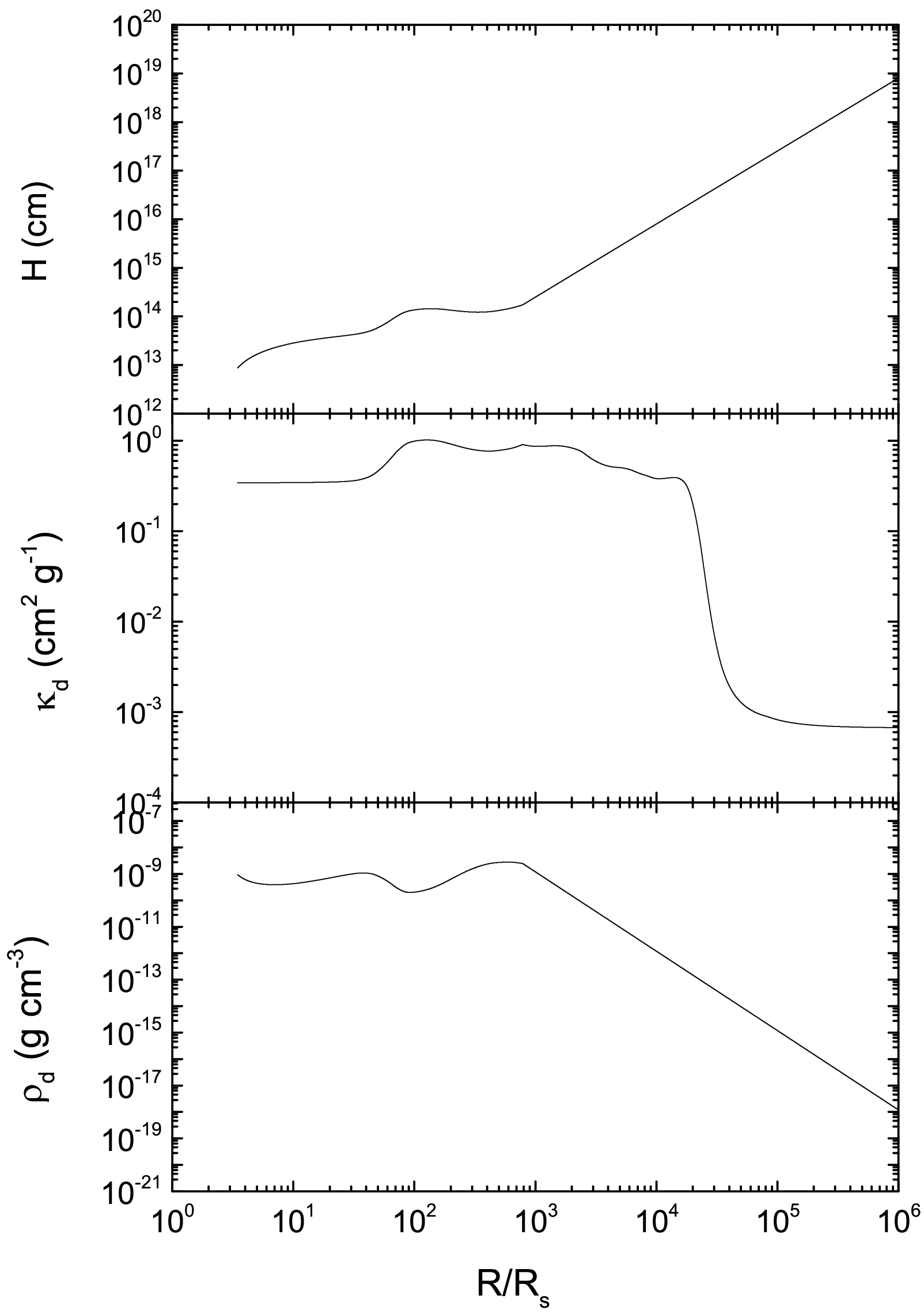}
\includegraphics [angle=0,scale=0.2] {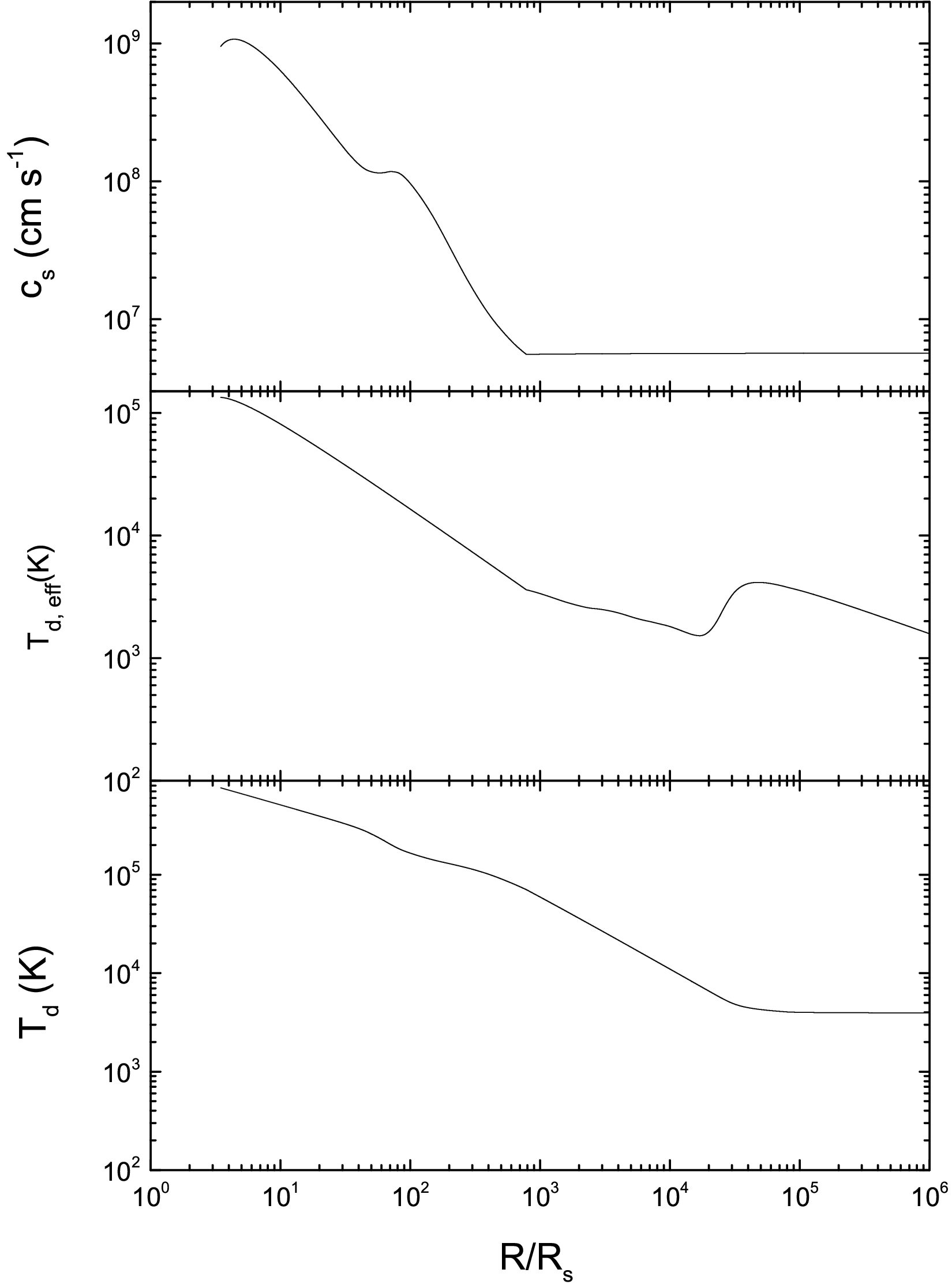}
\caption{The scale height of the AGN disk ($H$), opacity of the middle disk ($\kappa_{\rm d}$), the mid-plane density ($\rho_d$), the mid-plane sound velocity ($c_{\rm s}$), the effective temperature ($T_{\rm d,eff}$) and the mid-plane temperature ($T_{\rm d}$) of the AGN disk evolve with $R$. $\alpha=0.01$, $\epsilon_{\rm d}=0.1$, $M_{\rm SMBH}=10^8M_\odot$ and $l_{\rm E}=0.5$ are adopted. The inner and outer boundaries of the disk are set to $R_{\rm min}=R_{\rm s}/(4\epsilon_{\rm d})$ and $R_{\rm max}=10^6R_{\rm s}$, respectively.}
\label{fig:AGN}
\end{figure*}

For the outer disk, the influence of self-gravity of the AGN disk becomes more important. The stability is described with the Toomre’s parameter,
\begin{equation}
\label{eq:Toomre}
Q=\frac{c_{\rm s}\Omega_{\rm d}}{\pi G\Sigma_{\rm d}}\approx\frac{\Omega_{\rm d}^2}{2\pi G\rho_{\rm d}}.
\end{equation}
If $Q\leq 1$, the gas will become unstable due to the self-gravity and lead to the star formation in AGN disk. Following \citet{SirkoGoodman(2003)}, additional heating (such as nuclear fusion inside the star) is considered such that $Q=Q_{\rm min}\approx 1$. Then we can replace Eq.\ref{eq:T_d_eff} with
\begin{equation}
\label{eq:T_eff_1}
\rho_{\rm d}=\frac{\Omega_{\rm d}^2}{2\pi G Q_{\rm min}},
\end{equation}
for outer disk. By solving these equations, we can obtain the structure and properties of the inner and outer disk regions. The results are shown in Fig.\ref{fig:AGN}.

As we can see, the disk density is $R$-dependent. We need to determine the possible location of the GRBs $R_{\rm GRB}$ in the disk. Massive stars would migrate inward after forming in the outer region of the AGN disk \citep{LinPapaloizou(1993)}. \citet{Lietal.(2023)} demonstrate that the migration distance is relatively small and they studied the equilibrium radius of the wind of massive stars and the AGN disk material at different locations (e.g., $\sim 10^3$, $10^4$ and $10^5R_{\rm s}$) and found that the ratio of the cavity radius to the AGN disk height is only $\sim 0.001-0.01$. If this is the case for LGRB, the influence of the cavity on the dynamics of the jet in the AGN disk might not be significant. For $M_{\rm SMBH}=10^8 M_\odot$, \citet{Pernaetal.(2021)} shows that for a typical LGRB, the deceleration radius of the jet is much smaller than the photosphere radius of the AGN disk, regardless of its location in the AGN disk. Therefore, the choice of progenitor position has little effect on jet dynamics and radiation characteristics. In the following calculation, we adopt $\sim 1000 R_{\rm s}$ as a fiducial value.

For the SGRB progenitor, we should take into account the further migration of binary compact objects resulting from the collapse of massive stars. Such migration will terminate at migration traps $\sim 500 R_{\rm s}$, which would be the location for SGRBs in AGN disk \citep{Bellovaryetal.(2016)}.

Fig.\ref{fig:AGN} reveals that the AGN disk exhibits comparable magnitudes in terms of its height, density, and sound velocity at these two positions (i.e., $1000 R_{\rm s}$ for LGRBs and $500 R_{\rm s}$ for SGRBs). For simplicity, we assume that the massive stars or the binary compact objects are corotating with the AGN disk
and adopt $H\sim 10^{14}~ \rm cm$, $\rho_{\rm d}\sim 10^{-10}~\rm g~cm^{-3}$ and $c_{\rm s}\sim 5\times 10^6~ \rm cm~s^{-1}$ as typical values at the disk location of both short and long GRBs in the following calculations.

\subsection{The Cavity Induced by GRB Progenitor Wind}\label{Sec:jet structure}
As the progenitor of LGRBs, a massive star produce strong stellar wind from its surface \citep{Maeder(1983)}. For binary NSs or NS-BH binaries (as the progenitor of SGRBs), violent wind can be launched from the accretion disk during the super Eddington accretion process \citep{Jiangetal.(2014), Hashizumeetal.(2015)}.

Now, we consider the interactions between the GRB progenitor wind and the AGN disk gas, which will change the density distribution of the surrounding disk material. The wind would blow the material into a low-density region, called a cavity \citep{BranchWheeler(2017)}. We will study the cavity produced by the two types of GRB progenitor, respectively.

\subsubsection{Wind from SGRB progenitor}
First, we investigate the wind emanating from the SGRB progenitor, i.e., BCOs. We consider a binary composed of two compact stars of equal mass, that is, $m_1=m_2$, and total mass $m_{\rm b}=m_1+m_2\approx 5M_\odot$. They are intially orbiting around the SMBH at approximately $R_{\rm GRB}\sim 500R_{\rm s}$ in the AGN disk. The separation between the two compact stars is taken as $a_{\rm b}=10^{10} \rm cm$, below which the angular momentum loss in the binary is dominated by gravitational wave radiation \citep{Tagawaetal.(2020)}.

The BCOs system can be regarded as a point source when the binary separation $a_{\rm b}$ is much smaller than the outer boundary of the disk $r_{\rm out}$ \citep{Kimuraetal.(2021)}.
A circumbinary disk will form around the BCOs \citep{Kimuraetal.(2021)}. The viscosity parameter $\alpha_{\rm b}=0.1$ and the height ratio $h_{\rm b}=H_{\rm b}/r=0.5$ are adopted to model the circumbinary disk.

Based on Bondi–Holye–Lyttleton (BHL) model, \citet{Chenetal.(2023)} discussed the formation and evolution of the circumbinary disk in the AGN disk. The outer boundary of the circumbinary disk $r_{\rm out}$ can be estimated with the circularization radius $r_{\rm cir}$. Applying the principle of conservation of angular momentum to the infalling gas, one has $r_{\rm cir}\times \sqrt{\frac{Gm_{\rm b}}{r_{\rm cir}}}=r_{\rm gra}\times v_{\rm gra}$, where $r_{\rm gra}=\min\{r_{\rm BHL},r_{\rm Hill}\}$ is the radius of the BCOs gravity sphere, $v_{\rm gra}\approx 1/2 r_{\rm gra} \Omega_{\rm d}$ is the velocity of the gas at $r_{\rm gra}$, $r_{\rm Hill}=\left(m_{\rm b}/3M_{\rm SMBH}\right)^{1/3}R_{\rm GRB}$ is the Hill radius and $r_{\rm BHL}=G m_{\rm b}/c_{\rm s}^2$ is the BHL radius \citep{Edgar(2004)}. We find $r_{\rm out}\sim 8\times 10^{11}~\rm cm$.

The initial inflow mass rate at the outer boundary of the circumbinary disk $\dot{m}_{\rm out}$ is estimated with the BHL accretion rate and corrected for with the height of the AGN disk and the Hill radius,
\begin{equation}
\label{eq:M_dot_obd}
\dot{m}_{\rm out}=\dot{m}_{\rm BHL}\times {\rm min}\left\{1,\frac{H}{r_{\rm BHL}}\right\}\times {\rm min}\left\{1,\frac{r_{\rm Hill}}{r_{\rm BHL}}\right\},
\end{equation}
where $\dot{m}_{\rm BHL}=\frac{4\pi G^2 m_{\rm b}^2\rho_{\rm d}}{c_{\rm s}^3}$ is the BHL accretion rate.

We find $\dot{m}_{\rm out}\sim 6\times 10^6\dot{m}_{\rm Edd}$ with our parameter setup for the BCOs and AGN disk, and the Eddington accretion rate is $\dot{m}_{\rm Edd} =L_{\rm Edd,b}/(\epsilon_{\rm b} c^2)$, where $L_{\rm Edd,b}=4\pi Gm_{\rm b}m_{\rm p}c/\sigma_{\rm T}$ is the Eddington luminosity and $\epsilon_{\rm b}\sim 0.1$ is the radiation efficiency. Photons will be trapped in such super-Eddington accretion flow in the inner region. The trapping radius is \citep{Kitakietal.(2021)}
\begin{equation}
\label{eq:trap}
r_{\rm tr}=3h_{\rm b}\frac{\dot{m}_{\rm out}}{\dot{m}_{\rm Edd}}\frac{Gm_{\rm b}}{c^2}=3h_{\rm b}\frac{\dot{m}_{\rm out}}{\dot{m}_{\rm Edd}}r_{\rm g},
\end{equation}
where $r_{\rm g}=\frac{G m_{\rm b}}{c^2}$ is the gravitational radius of the BCOs. For $m_{\rm b}=5M_{\rm \odot}$, $r_{\rm tr}\sim 7\times 10^{12}~\rm cm$.

The supper-Eddington accretion disk will launch in a strong wind due to the influence of radiation pressure. When $r_{\rm out}<r_{\rm tr}$,  the mass accretion rate onto BCOs will deviate from that at the outer boundary \citep{Yangetal.(2014)},
\begin{equation}
\label{eq: M_dot_acc}
\dot{m}_{\rm acc}=\dot{m}_{\rm out} (r/r_{\rm out})^s,
\end{equation}
The parameter $s$ describes the effects of wind, which varies from 0 to 1 \citep{Yangetal.(2014)}.
We adopt $s=0.06$ as a fiducial value in the calculations \citep{Kimuraetal.(2021),YuanChenChaoetal.(2022)} \footnote{In \citet{Kimuraetal.(2021)} and \cite{YuanChenChaoetal.(2022)}, a parameter $\eta_{\rm w}= \dot{m}_{\rm w}/\dot{m}_{\rm out}$ is introduced to describe the outflow rate, and $\eta_{\rm w} \sim 0.3$ is adopted as a fiducial value. Here, $s=0.06$ corresponds to $\eta_{\rm w} \sim 0.3$.}.
We define $\dot{m}_{\rm out,w}$ as the mass accretion rate at $r_{\rm out,w}$ of the circumbinary disk, $r_{\rm in,w}$ and $r_{\rm out,w}$ are the inner and outer boundaries of the wind launching region, respectively. We adopt $r_{\rm in,w}\sim C_{\rm gap} a_{\rm b}$ as the inner boundary of the circum-binary disk\citep{Nixon2013, Kimuraetal.(2021),YuanChenChaoetal.(2022), Leeetal.(2024)}, where $C_{\rm gap}\sim 2$ is the parameter to measure the gap size between the circum-binary disk and the binary and $a_{\rm b}$ is the binary major axis \footnote{Inside the inner boundary of the circum-binary disk, the accretion gas forms tow mini-disks surrounding each compact star. The mini-disks also can produce the outflow \citep{Kimuraetal.(2021),YuanChenChaoetal.(2022)}. For simplicity, we assume the density profile will not be significantly changed by the mini-disk outflow.}.

The outer boundary of the wind launching region can be estimated as $r_{\rm out,w}={\rm min}\{r_{\rm out}, r_{\rm tr}\} \sim r_{\rm out}$, then $\dot{m}_{\rm out,w}=\dot{m}_{\rm out} (r_{\rm out,w}/r_{\rm out})^s\approx \dot{m}_{\rm out}$
The density of the wind can be described as a spherical distribution \citep{Kimuraetal.(2021)},
\begin{equation}
\label{eq:rho_w}
\rho_{\rm b,w}\approx \frac{\dot{m}_{\rm w}}{4\pi r^2 v_{\rm b,w}}\approx
7.9\times 10^{-7} ~\dot{m}_{\rm w,24}r_{10}^{-2} v_{b,w,9}^{-1}~~\rm g~cm^{-3},
\end{equation}
where $\dot{m}_{\rm w}=\dot{m}_{\rm out,w}-\dot{m}_{\rm in,w}$ is the outflow rate of the disk wind, $\dot{m}_{\rm in,w}$ is the accretion rate at $r_{\rm in,w}$, and
$v_{\rm b,w,9}=v_{\rm b,w}/10^9 ~\rm cm~s^{-1}$ is the wind velocity \citep{Kimuraetal.(2021),Yuanetal.(2022)}, $\dot{m}_{\rm w,24}=\dot{m}_{\rm w}/10^{24}~\rm g~s^{-1}$ and $r_{10}=r/10^{10}~\rm cm$.

The wind will expand consistently outwards in the accretion timescale $t_{\rm acc}$, leading to a cavity in the AGN disk. As a result, the cavity will terminate the accretion, and the material from AGN disk will refill the cavity subsequently in a timescale $t_{\rm ref}$. The accretion timescale $t_{\rm acc}$ can be estimated with the viscous timescale $t_{\rm vis}$ at $r_{\rm out}$,
\begin{equation}
\label{eq:t_acc}
t_{\rm acc}\approx t_{\rm vis} (r_{\rm out})=1/(\alpha_{\rm b}h_{\rm b}^2\Omega_{\rm K,b}),
\end{equation}
where $\Omega_{\rm K,b}=\sqrt{Gm_{\rm b}/r^3}$. For $m_{\rm b}=5M_\odot$, the accretion timescale $t_{\rm acc}\sim 10^6\rm s$.

The interaction between the wind and the material of AGN disk would form a shocked shell.
The evolution of the shock shell's radius $r_{\rm b,sh}$ and velocity $v_{\rm b,sh}$ are \citep{Weaveretal.(1977)},
\begin{equation}
\label{eq:r_w}
r_{\rm b,sh}=0.88\left(\frac{L_{\rm b,w} t^3}{\rho_{\rm d}}\right)^{1/5},
\end{equation}
\begin{equation}
\label{eq:v_w}
v_{\rm b,sh}=0.35\left(\frac{L_{\rm b,w}}{\rho_{\rm d} t^2}\right)^{1/5}.
\end{equation}
where $L_{\rm b,w}=\dot{m}_{\rm w}v_{\rm b,w}^2\approx 1\times 10^{42}\dot{m}_{\rm w,24}v_{\rm b, w,9}^2\rm ~erg~s^{-1}$ is the kinetic luminosity
of the outflows.
Now, we can investigate the possibility of SGRB progenitor wind penetrating from the AGN disk surface during the BCOs accretion timescale. By equating $r_{\rm b,sh}\sim H$, we can determine the penetrating timescale $t_{\rm bre}\sim t_{\rm acc}$.
The binary merger timescale due to gravitational wave radiation is denoted as\citep{ShapiroTeukolsky(1983)}, $t_{\rm merge}=\frac{5}{128}\frac{c^5 a_b^4}{G^3 m_b^3}\sim 3\times 10^9 {\rm s} \gg t_{\rm acc}$. Therefore, the wind can effectively penetrate the surface of the AGN disk prior to merger.

The shell will continue to propagate in the low-density region after it penetrates the AGN disk vertically ($t_{\rm bre}$). Meanwhile, in the horizontal direction, the wind continues to interact with the AGN disk material. After the penetration, the evolution of the radius($r_{\rm b,shb}$) and velocity ($v_{\rm b,shb}$) of the shell satisfies\citep{Chenetal.(2023)},
\begin{equation}
\label{eq:r_wb}
r_{\rm b,shb}=r_{\rm b,sh}(t_{\rm bre})(t/t_{\rm bre})^{1/2},
\end{equation}
\begin{equation}
\label{eq:v_wb}
v_{\rm b,shb}=v_{\rm b,sh}(t_{\rm bre})(t/t_{\rm bre})^{-1/2}.
\end{equation}
The shell enters snowplow phase as the BCOs accretion stops. The propagation stops when the velocity of the shell approaches the speed of sound, and the radius of the wind (or cavity, $r_{\rm b,cav}$) reaches the maximum. This can be derived by conservation of momentum\citep{Chenetal.(2023)},
\begin{equation}
\label{eq:r_cav}
r_{\rm b,cav}=\left[\frac{r_{\rm b,shb}^2(t_{\rm acc})v_{\rm b,shb}(t_{\rm acc})}{c_{\rm s}}\right]^{1/2}.
\end{equation}
combining with the Eqs.(\ref{eq:t_acc}), (\ref{eq:r_wb}) and (\ref{eq:v_wb}), one can derive $r_{\rm b,cav}\sim 3\times 10^{14}~\rm cm$.
The time scale for the cavity to be refilled by the dense gas in the AGN disk is approximately \citep{Wangetal.(2021)},
\begin{equation}
\label{eq:t_ref}
t_{\rm ref}=r_{\rm b,cav}/c_{\rm s}.
\end{equation}
according to Eq.(\ref{eq:r_cav}), $t_{\rm ref}\approx 1.2\times 10^7~s\sim 50t_{\rm acc}$
The merger of BCOs can occur during either the accretion process or refill stage. The filled stage resembles the case discussed in \citet{Zhuetal.(2021)}, and is referred to as ``no-cavity'' in this work.
In the subsequent calculations, we mainly focus on the penetrating case ($r_{\rm b,cav}\sim H$) for BCOs, and refer to it as ``SGRB''. The density profile is thus given by $\rho=\rho_{\rm b,w}$ for $r>r_{\rm in, w}$.

\subsubsection{Wind from LGRB progenitor}

The progenitor of LGRBs is likley a Wolf-Rayet (WR) star  \citep{WoosleyBloom(2006)}. The typical radius and effective temperature of WR stars are $r_{\rm *}=5R_\odot$ and $T_{\rm *}=10^5~\rm K$, respectively \citep{Crowther(2007), SchaererMaeder(1992)}.

The wind luminosity can be written as,
\begin{equation}
\label{eq:L_*_1}
L_{\rm *,w}=\dot{m}_{\rm *} v_{\rm \infty} c,
\end{equation}
where $\dot{m}_{\rm *}$ is the mass loss rate of the star and $v_\infty$ is the terminal velocity of the wind. For WR stars, $v_{\rm \infty} \sim 1000~\rm km/s$ \citep{Crowther(2007)}. The primary source of the stellar wind luminosity is the thermal radiation from the stellar surface. Therefore, we have
\begin{equation}
\label{eq:L_*_2}
L_{\rm *,w}\approx 4\pi r_{*}^2\sigma_{\rm SB}T_{\rm *}^4 .
\end{equation}
According to Eqs. (\ref{eq:L_*_1}) and (\ref{eq:L_*_2}), one can derive that
\begin{equation}
\label{eq:m_*_dot}
\dot{m}_* =\frac{4\pi r_*^2 \sigma_{\rm SB}T_*^4}{v_\infty c}.
\end{equation}
The density distribution of the stellar wind $\rho_{\rm *,w}$ is
\begin{equation}
\label{eq:rho_*_w}
\rho_{\rm *,w}\approx \frac{\dot{m}_{\rm *}}{4\pi r^2 v_{\rm\infty}} .
\end{equation}

The wind will also develop a cavity in the AGN disk. The cavity size can be determined by considering the balance between the ram pressure of the stellar wind $L_{\rm *,w}/4\pi r^2 c$ and the thermal pressure of the AGN disk material $P_{\rm d} \sim 2\times 10^3 ~{\rm erg~cm^{-3}}$, i.e.,
\begin{equation}
\label{wind_shell}
r_{\rm *,cav}=\sqrt{\frac{L_{\rm *,w}}{4\pi cP_{\rm d}}},
\end{equation}
which is $\sim0.05H$ for the parameters setting in this paper, and such case is referred to as ``LGRB''.

The reconstructed density profile surrounding the GRBs in AGN disk can be described with a piece-wise function for LGRB case,
\begin{equation}
\label{eq: rho_r}
\rho=\left\{
\begin{aligned}
&\rho_{\rm *,w}, &r_{\rm *}<r<r_{\rm *,cav} \\
&\rho_{\rm AGN}, &r>r_{\rm *,cav}
\end{aligned}
\right. ,
\end{equation}
where $\rho_{\rm AGN}=\rho_{\rm d} \times \exp(-\frac{r^2}{2H^2})$ is the density distribution of the AGN disk in the vertical direction \citep{Kathirgamarajuetal.(2023)}.

\begin{figure}
\centering
\includegraphics [angle=0,scale=0.3] {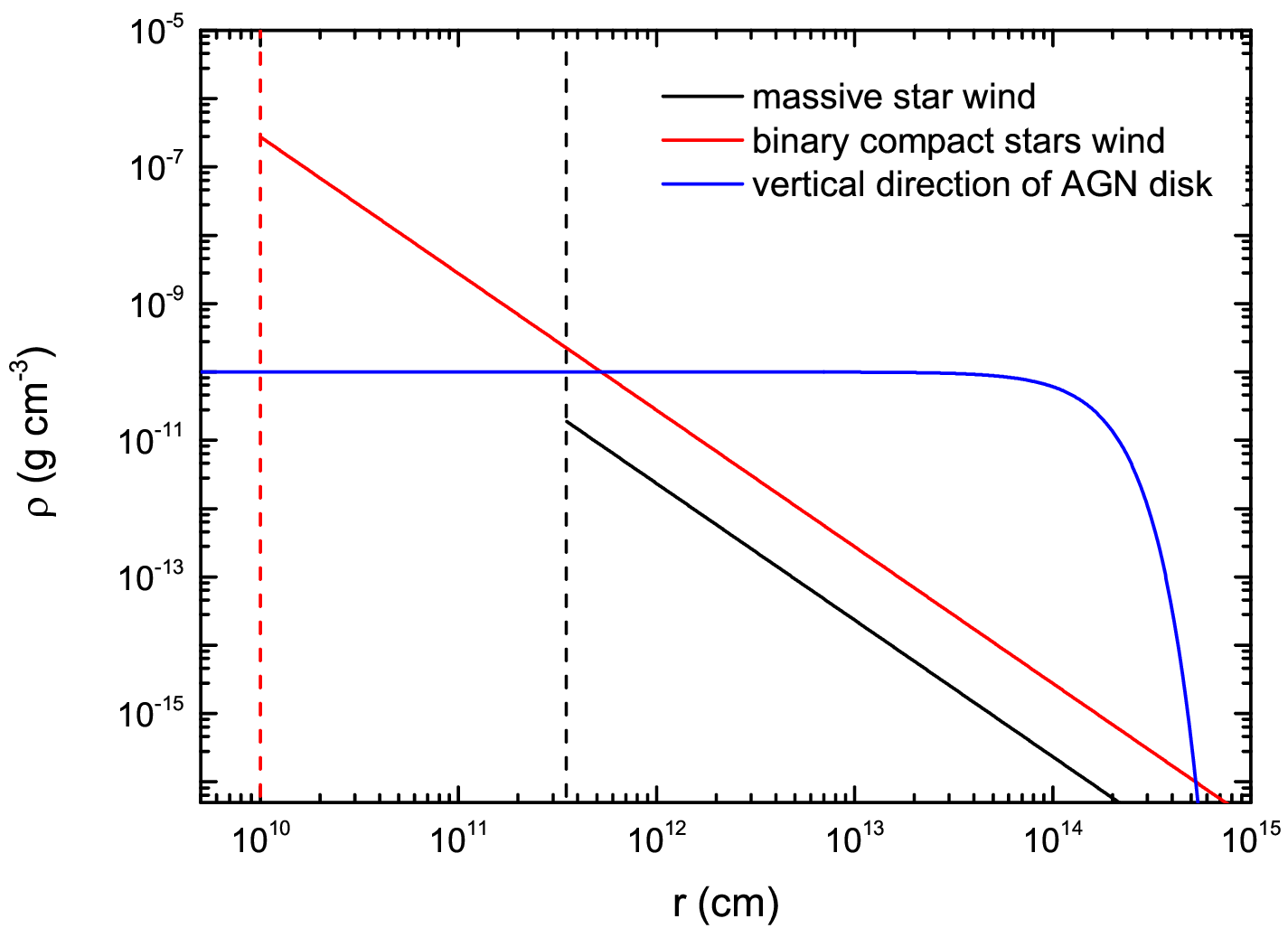}
\caption{The density distribution of the LGRB progenitor (massive star) wind (black), the SGRB progenitor (BCOs) wind (red), and the AGN disk (blue) in vertical direction. The red and black dashed lines represent the inner boundaries of the BCOs wind and the massive star wind, respectively.}
\label{fig:rho_R}
\end{figure}

Fig.\ref{fig:rho_R} presents the density profile for AGN disk (blue), BCO wind (red) and stellar wind (black). The vertical dashed lines represent the inner boundaries of progenitor wind.

\section{GRB with Two-component Jet Propagating in the AGN Disk} \label{disk GRB}
\begin{figure}
\centering
\includegraphics [angle=0,scale=0.33] {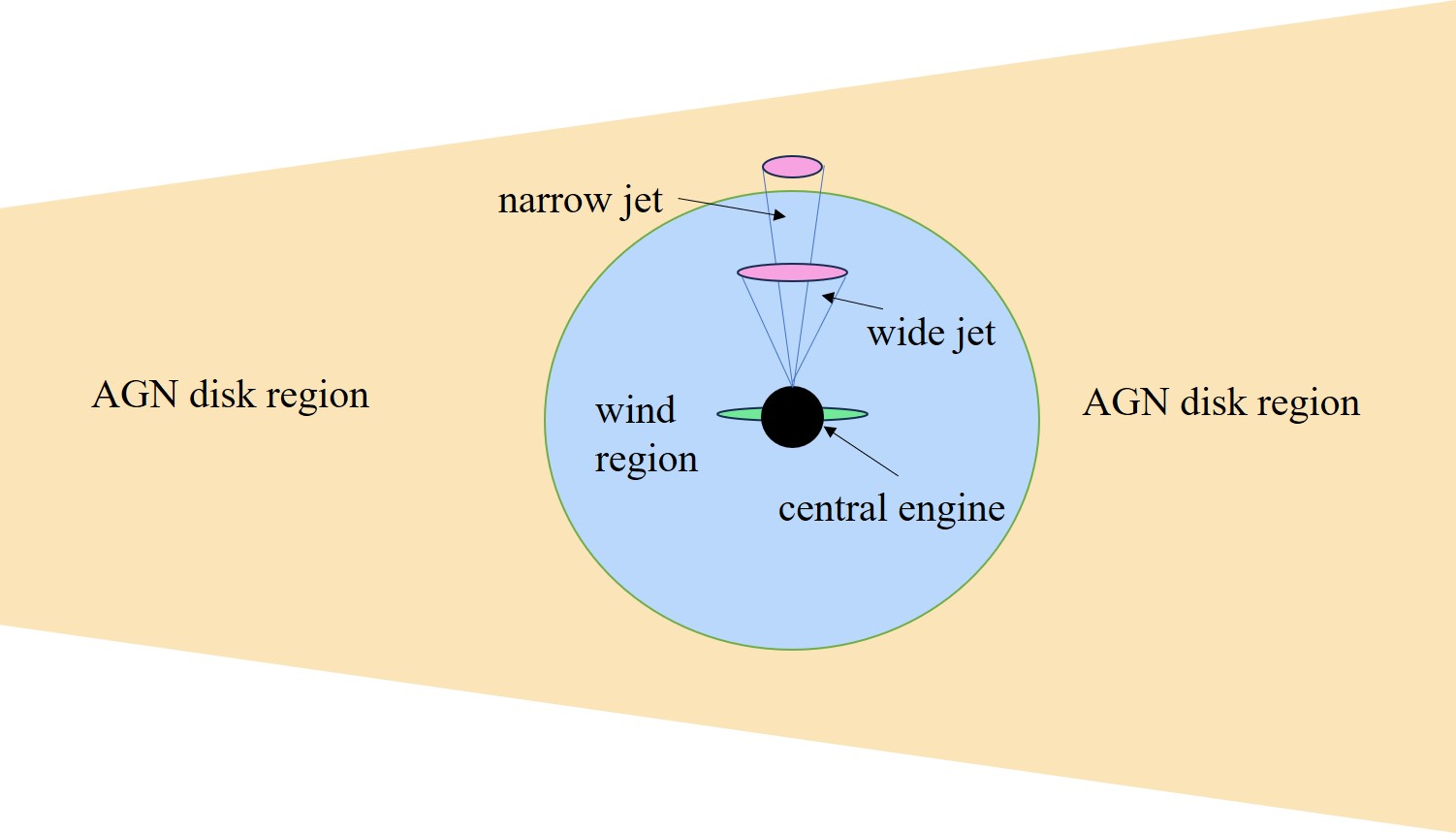}
\caption{This schematic illustrates the structure of a jet and its surrounding environment within an AGN disk. The orange shaded region represents the AGN disk material, while the light blue region denotes the cavity developed by the GRB progenitor wind. We assume a GRB with a two components jet launched in mid-plane of the disk.}
\label{fig:sketch}
\end{figure}

Now, we consider a GRB with a two-component jet that occurs in the accretion disk environment of an AGN with cavity, as shown in Fig. \ref{fig:sketch}.

\subsection{Jet Dynamic in AGN Disk with Cavity}\label{Dynamic Equations}

We consider a shell with isotropic energy $E_{0, \rm iso}$, initial Lorentz factor $\Gamma_0$, and half-opening angle $\theta_{\rm j0}$ launched from the midplane of the AGN disk and directed perpendicularly to the disk plane, propagating into the ambient medium described in Section \ref{Cavity}. The number density can generally be described as $n=Ar^{-k}$, where $k$ is the power-law index (e.g., $k=0$ for the AGN disk environment and $k=2$ for the wind ambient).
The interaction between the shell and the medium is described by two shocks: a reverse shock (RS) propagating into the shell and a forward shock (FS) propagating into the ambient medium \citep{ReesMeszaros(2005), Piran(1999), SariPiran(1995),Yuan2025}. The two shocks divide the ambient medium and shell into four regions \citep{SariPiran(1995)}: (1) the unshocked ambient medium region; (2) the shocked ambient medium region; (3) the shocked shell material region; (4) the unshocked shell material. First, the pair of shocks (FS and RS) propagate into the medium and the shell, respectively. After the RS crosses the shell, the blast wave enters the deceleration phase. After the blastwave has sufficiently decelerated, the shell enters the post-jet-break phase when the $1/\Gamma$ cone exceeds $\theta_{\rm j}$. The blastwave finally reaches the Newtonian phase the blast wave enters the Newtonian phase when it has swept up the ISM with the total rest mass energy comparable to the energy of the ejecta.

We introduce a timescale over which the accumulated ambient medium mass is $1/\Gamma_0$ of the ejecta mass, i.e., $t_{\rm dec}=\left[(3-k)E_{\rm 0,iso}/(2^{4-k}\pi Am_{\rm p}\Gamma_0^{8-2k}c^{5-k}\right]^{1/(3-k)}$ \citep{Zhangetal.(2003), Meszarosetal.(1998), Kobayashi(2000)}. If
the burst duration $t_{\rm j}<t_{\rm dec}$, i.e., the thin shell scenario, this timescale defines the jet deceleration time. If $t_{\rm j}>t_{\rm dec}$, i.e., the thick shell scenario, the deceleration time is delayed to $t_{\rm j}$.

The time of RS crossing the shell  can be defined as $t_{\rm c}=\max (t_{\rm dec}, t_{\rm j})$. For GRBs in the dense medium we studied in this paper, we have $t_{\rm dec}< t_{\rm j}$ for LGRB and SGRB. Therefore, we adopt the thick shell scenario and $t_{\rm c}=t_{\rm j}$ in the calculation.

Before RS crossing the shell, the evolution of the RS can be scaling as \citep[see][for a review]{Yietal.(2013),GaoMeszaros(2015)},
\begin{eqnarray}
\label{eq_rs_thick_before}
\gamma_3 = \gamma_{\rm 3,c} (t/t_{\rm c})^{\frac{-(2-k)}{2(4-k)}}, \ n_3 = n_{\rm 3,c} (t/t_{\rm c})^{\frac{-(6+k)}{2(4-k)}},\\ \nonumber
e_3 = e_{\rm 3,c} (t/t_{\rm c})^{\frac{-(2+k)}{(4-k)}}, ~~~~~~~~~~~~~\ N_{\rm 3} = N_{\rm 3,c}~ t/t_{\rm c}.
\end{eqnarray}

The evolution of the FS before RS crossing can be scaling as \citep[see][for a review]{Yietal.(2013),GaoMeszaros(2015)},
\begin{eqnarray}
\label{eq_fs_thick_before}
\gamma_2 = \gamma_{\rm 2,c} (t/t_{\rm c})^{\frac{-(2-k)}{2(4-k)}}, \ n_2 = n_{\rm 2,c} (t/t_{\rm c})^{\frac{-(2+3k)}{2(4-k)}},~~~~\\ \nonumber
e_2 = e_{\rm 2,c} (t/t_{\rm c})^{\frac{-(2+k)}{(4-k)}}, ~~~~~~~~~~~~~\ N_{\rm 2} = N_{\rm 2,c} (t/t_{\rm c})^{\frac{2(3-k)}{(4-k)}},
\end{eqnarray}
where $\gamma_i$, $n_i$, $e_i$ and $N_{\rm e,i}$ are the Lorentz factor, comoving number density, energy density and the number of electrons of Region (i). $N_0=M_{\rm ej}/m_{\rm p}$ is the total number of electrons in the shell.
$\gamma_{\rm i,c}$, $n_{\rm i,c}$, $e_{\rm i,c}$ and $N_{\rm i,c}$ are the Lorentz factor, comoving number density, energy density and the number of electrons at the crossing time $t_{\rm c}$. $\gamma_{\rm 3,c}=\gamma_{\rm 2,c}\simeq \Gamma_0^{1/2}f_{\rm c}^{1/4}$, where $f_{\rm c}=\frac{l^{3-k}}{(3-k)r_{\rm c}^{2-k}ct_{\rm c}\Gamma_0^2}$ is the density ratio of region 4 and 1, $l=\left(\frac{(3-k)E_{\rm 0,iso}}{4\pi Am_{\rm p}c^2}\right)^{1/(3-k)}$ is the Sedov length, and $r_{\rm c}=\left(\frac{(4-k)^2l^{3-k}ct_{\rm c}}{16(3-k)}\right)^{1/(4-k)}$ is the crossing radius. $n_{\rm 3,c}=\frac{8\gamma_{\rm 3,c}n_{1,c}}{\Gamma_0}$ and $e_{\rm 3,c}=\gamma_{\rm 34,c}n_{\rm 3,c}m_{\rm p}c^2$, where $\gamma_{\rm 34,c}=\Gamma_0/(2\gamma_{\rm 3,c})$ is the the relative Lorentz factor between the region 3 and 4 at $t_{\rm c}$. $n_{\rm 2,c}=4\gamma_{\rm 2,c}n_{\rm 1,c}$ and $e_{\rm 2,c}=\gamma_{\rm 2,c}n_{\rm 2,c}m_{\rm p}c^2$. $N_{\rm 3,c}=N_0$ and $N_{\rm 2,c}=4\pi n_{\rm 1,c}r_{\rm c}^3/(3-k)$.

After the RS crossing the shell, the evolution of the RS can be described as,
\begin{eqnarray}
\label{eq_rs_thick_after}
\gamma_3 = \gamma_{\rm 3,c} (t/t_{\rm c})^{\frac{(2k-7)}{4(4-k)}}, \ n_3 = n_{\rm 3,c} (t/t_{\rm c})^{\frac{(2k-13)}{4(4-k)}},\\ \nonumber
e_3 = e_{\rm 3,c} (t/t_{\rm c})^{\frac{(2k-13)}{3(4-k)}},\ N_{\rm e,3} \simeq N_0.
\end{eqnarray}

The dynamical evolution of the FS after the RS crossing is calculated numerically using the equations \citep{Huangetal.(2000),Pe'er(2012)},
\begin{equation}
\label{Eq:dynamic_1}
\frac{dr}{dt}=\beta c \Gamma \left(\Gamma + \sqrt{\Gamma^2-1}\right),
\end{equation}

\begin{equation}
\label{Eq:dynamic_2}
\frac{dm}{dr}=2\pi r^2 \left(1-cos\theta_{\rm j}\right)n_1 m_{\rm p},
\end{equation}

\begin{equation}
\label{Eq:dynamic_3}
\frac{d\theta_{\rm j}}{dt}=\frac{c_{\rm s,j}\left(\Gamma+\sqrt{\Gamma^2-1}\right)}{r},
\end{equation}

\begin{equation}
\label{Eq:dynamic_7}
\frac{d\Gamma}{dm}=-\frac{\hat{\gamma}(\Gamma^2-1)-(\hat{\gamma}-1)\Gamma\beta^2}
{m_{\rm ej}+\epsilon m+(1-\epsilon)m[2\hat{\gamma}\Gamma-(\hat{\gamma}-1)(1+\Gamma^{-2})]},
\end{equation}
where $r$ and $t$ are the radius and time of the jet in the source frame, $\beta=\sqrt{\Gamma^2-1}/\Gamma$ is the dimensionless speed of the jet, $m$ is the swept-up mass, $n_1=\rho/m_{\rm p}$ is the medium density given by Eq. (\ref{eq: rho_r}). $m_{\rm ej}=E_{\rm j}/(\Gamma_0 c^2)$ is ejecta mass, and $E_{\rm j}=E_{\rm 0,iso}(1-\cos\theta_{\rm j0})$ is the kinetic energy of the ejecta. $\epsilon$ is the radiative efficiency of the jet, and $\epsilon = 0$ (adiabatic scenario) is adopted throughout this paper. Eq. (\ref{Eq:dynamic_3}) describes the side-way expanding of jet in the medium,
 where $c_{\rm s,j}$ is the sound speed of the shocked ambient medium which can be read as \citep{KirkDuffy(1999)},
\begin{equation}
\label{Eq:dynamic_4}
c_{\rm s,j}^2=\frac{\hat{\gamma}\left(\hat{\gamma}-1\right)\left(\Gamma-1\right)}
{1+\hat{\gamma}\left(\Gamma-1\right)}c^2,
\end{equation}
where $\hat{\gamma}$ is the adiabatic index. For simplicity,, we adopt a polynomial fit formal to calculate the adiabatic index evolution the shocked ambient medium \citep{Service(1986)}, $\hat{\gamma}=(5-1.21937z+0.18203z^2-0.96583z^3+2.32513z^4 -2.39332z^5+1.07136z^6)/3$,
where $z=\Theta/(0.24+\Theta)$ and $\Theta\simeq \frac{\Gamma\beta}{3}
\frac{\Gamma\beta+1.07(\Gamma\beta)^2}{1+\Gamma\beta+1.07(\Gamma\beta)^2}$.

\citet{Zhuetal.(2021)} pointed out that the GRB jet will be choked in the AGN disk. However, they did not consider the reconstruction of density profile due to the GRB progenitor wind \citep{Yuanetal.(2022)}. Therefore, we need to check whether the GRB jet can successfully break out from the AGN disk with cavity.

The jet could be choked if its tail catches up to the head. This occurs if $(v_{\rm t}-v_{\rm h})t_{\rm j,bo}>z_{\rm j}$, or
\begin{equation}
\label{Eq: t_j_bo}
t_{\rm j,bo}>2\Gamma^2t_{\rm j},
\end{equation}
where $v_{\rm t}$ and $v_{\rm h}$ are the tail and head velocity of the jet, $t_{\rm j,bo}$ is the timescale that the jet breakout from the AGN disk and $z_{\rm j}\approx ct_{\rm j}$ is the length of the jet and $t_{\rm j}$ is the duration of the GRB in the burst frame.
The jet choked time is $t_{\rm ch}=z_{\rm j}/(v_{\rm t}-v_{\rm h})=t_{\rm j}/(1-\beta_{\rm h})\approx 2\Gamma^2(t_{\rm ch})t_{\rm j}$ \citep{Zhuetal.(2021),Zhangetal.(2024)}.

The choked jet will then deposit its energy within the disk materials to form a hot cocoon with energy $E_{\rm c}\approx E_{\rm j}$.
The cocoon undergoes expansion, leading to the formation of a radiation-mediated shock that sweeps through the AGN disk material until it ultimately breaks out.

\subsection{Emission from Forward and Reverse Shocks}\label{Sec:FSRSEmission}

During the dynamical evolution of FS and RS, electrons are thought to be accelerated from the shock front by the first order Fermi acceleration mechanism to power-law distribution, i.e., $N(\gamma_{\rm e})d\gamma_{\rm e} \propto \gamma_{\rm e}^{-p}d\gamma_{\rm e}$ ($\gamma_{\rm m}<\gamma_{\rm e}<\gamma_{\rm M})$, where $\gamma_{\rm e}$ is the Lorentz factor of the electrons and $\gamma_{\rm m}$ and $\gamma_{\rm M}$ are the minimum and maximum Lorentz factors of the injected electron \citep{Sari(1998), Meszarosetal.(1998)},
\begin{equation}
\label{eq_8}
\gamma_{\rm m}=\frac{g(p)\epsilon_{\rm e} (\Gamma -1)m_{\rm p}}{m_{\rm e}},
\end{equation}
where $\epsilon_{\rm e}$ is the fraction of the shock energy into electrons, and $m_{\rm e}$ is the mass of electron. The function $g(p)=(p-2)/(p-1)$ for $p>2$.
The maximum electron Lorentz factor $\gamma_M$ can be calculated by balancing the acceleration and dynamical time scales,
\begin{equation}
\label{eq_10}
\gamma_{\rm M}\simeq \frac{\Gamma q_{\rm e} B}{m_{\rm e} c}t,
\end{equation}
where $q_{\rm e}$ is the electron charge, and $B$ is the comoving magnetic field strength,
\begin{equation}
\label{eq_11}
B=\sqrt{8\pi e \epsilon_{\rm B}},
\end{equation}
where $e$ and $\epsilon_{\rm B}$ are the energy density in the shocked region and fraction of the shock energy in magnetic field.

The observed radiation power for synchrotron radiation of a single electron is \citep{RybickiLightman(1979)},
\begin{equation}
\label{eq_12}
P(\gamma_{\rm e})=\frac{4}{3}\sigma_T c \gamma_{\rm e}^2\frac{\Gamma^2B^2}{8\pi}.
\end{equation}
The characteristic frequency for an electron with Lorentz factor $\gamma_{\rm e}$ is,
\begin{equation}
\label{eq_13}
\nu(\gamma_{\rm e})=\Gamma \gamma_{\rm e}^2\frac{q_eB}{2\pi m_{\rm e} c}.
\end{equation}
The peak power occurs at $\nu(\gamma_e)$,
\begin{equation}
\label{eq_14}
P_{\nu,\rm max}=\frac{P(\gamma_{\rm e})}{\nu(\gamma_{\rm e})}=\frac{m_{\rm e} c^2\sigma_{\rm
 T}\Gamma B}{3q_{\rm e}}.
\end{equation}
The cooling Lorentz factor $\gamma_{\rm c}$ can be defined by equating the life time of a relativistic electron to the time $t$,
\begin{equation}
\label{eq: gamma_c}
\gamma_{\rm c}=\frac{6\pi m_e c}{\Gamma \sigma_{\rm T} B^2 t},
\end{equation}
While $\gamma_e > \gamma_c$, the cooling of the electrons would be significant, and further change the energy distribution of electrons.
In synchrotron radiation, in addition to the characteristic frequencies $\nu_m$ and $\nu_c$, there is an important characteristic frequency, self-absorption frequency $\nu_a$, which can be calculated by equating the synchrotron flux with the flux of a blackbody \citep{SariPiran(1999), KobayashiZhang(2003)}, or by the condition that the optical depth for self-absorption is unity \citep{RybickiLightman(1979)}.

The emissions from FS and RS are calculated by employing a modified version of the numerical code \texttt{PyFRS}\footnote{\url{https://github.com/leiwh/PyFRS}}  \citep{Gao+2013, Wang2014, Lei2016,Zhang2018, Zhu2023,Zhou2024}.

\subsection{Cocoon shock breakout Emission}\label{Sec: cocoon}

The cocoon shock breaks out when the dynamic timescale $t_{\rm dyn,c}=d_{\rm c}/v_{\rm c}$ and diffusion timescale $t_{\rm diff,c}=d_{\rm c}/c\int^{r_{\rm ph}}_{r_{\rm ph}-d_{\rm c}} \kappa_{\rm w}\rho dR$ of the cocoon are equal, where $r_{\rm
ph}$ is the photosphere radius of the ambient environment defined by $\int_{r_{\rm ph}}^\infty\kappa_{\rm w}\rho dr=1$, and $d_{\rm c}$ is the distance between the position of shock breakout and the photosphere radius. $\kappa_{\rm w}\approx 0.4~\rm cm^2~g^{-1}$ is the opacity of the ambient medium.
The mass of the breakout layer is $m_{\rm c,bo}\approx \pi\rho \theta_{\rm c}^2r_{\rm ph}^2d_{\rm c}$, where $\theta_{\rm c}$ is the opening angle of the cocoon.

The mass of the swept material by the cocoon in the position $r$ is,
\begin{equation}
\label{Eq: M_c}
m_{\rm c}=\int^{r}_{r_{\rm 0,w}}\pi\rho\theta_{\rm c}^2r^2dr.
\end{equation}
where $r_{\rm 0,w}$ is the inner boundary of the medium, e.g., $r_{\rm 0,w}=r_{\rm in,w}$ for SGRB case and $r_{\rm 0,w}=r_{\rm *}$ for LGRB case.
The velocity of the cocoon is $v_{\rm c}/c\approx \sqrt{1-1/\Gamma_{\rm c}^2}$, where $\Gamma_{\rm c}= E_{\rm c}/m_{\rm c}c^2+1$.
The time evolution of the cocoon opening angle $\theta_{\rm c}$ can be described as \citep{Irwinetal.(2019)},
\begin{equation}
\label{eq: Ttheta_c}
\theta_{\rm c}\sim \left\{
\begin{aligned}
&b_{\rm c}/a_{\rm c}, &t<t_{\rm b}+t_{\rm ch} \\
&(t/t_{\rm a})^{1/2}, &t_{\rm b}+t_{\rm ch}<t<t_{\rm a}+t_{\rm ch}\\
&1, &t>t_{\rm a}+t_{\rm ch}
\end{aligned}
\right.
\end{equation}
where $a_{\rm c}$ and $b_{\rm c}$ present the position  and the width of the cocoon when the jet choking. We assume $b_{\rm c}\ll a_{\rm c}$ (e.g., $b_{\rm c}\sim 0.1a_{\rm c}$, see \citet{Irwinetal.(2019)}), due to the velocity of the jet head is higher than the sideways expansion before the jet choked. $t_{\rm a}\sim (48\pi \rho (a_{\rm c})a_{\rm c}^5/7E_{\rm c} )^{1/2}$ is the timescale for the initial height of the cocoon to double, $t_{\rm b}\sim t_{\rm ch}$ is the timescale for the initial width of the cocoon to double \citep{Irwinetal.(2019)}.

Therefore, the velocity of the cocoon when the shock breaks out is, $v_{\rm c,bo}/c\approx \sqrt{1-1/\Gamma_{\rm c,bo}^2}$ and $\Gamma_{\rm c,bo}= E_{\rm c}/m_{\rm c}(r=r_{\rm ph}) c^2+1$.

The energy released by the layer that the shock breaking out is $E_{\rm c,bo}=m_{\rm c,bo}c^2(\Gamma_{\rm c,bo}-1)$, and the luminosity is,
\begin{equation}
\label{Eq: L_c_bo}
L_{\rm c,bo}=E_{\rm c,bo}/t_{\rm diff,c}.
\end{equation}
The duration of the shock breakout can be estimated as $t_{\rm bo}\approx t_{\rm dyn,c}$.

We calculate the radiation temperature according to the method described in \citep{NakarSari(2010), ChenDai(2023)}.
The thermal coupling coefficient of Compton equilibrium is defined as,
\begin{equation}
\label{Eq: eta}
\eta=\frac{n_{\rm BB}}{t_{\rm diff,c}\dot{n}_{\rm ph,ff}(T_{\rm BB})},
\end{equation}
where $T_{\rm BB}\approx \left[(\Gamma_{\rm c,bo}-1)\rho c^2/a\right]^{1/4}$ is the thermal equilibrium temperature of the downstream shock, and $a$ is the radiation constant. $n_{\rm BB}\approx aT_{\rm BB}^4/(3k_{\rm B}T_{\rm BB})$ is the number density of generated photons and $\dot{n}_{\rm ph,ff}\approx 3.5\times 10^{36}~s^{-1}~cm^{-3}~\rho^2 T^{-1/2}$. When $\eta <1$, the radiation can reach thermal equilibrium with $T_{\rm BB}$. When $\eta >1$, the radiation would deviate from the thermal equilibrium, forming a Compton equilibrium, and the Compton correction factor can be expressed as
\begin{equation}
\label{Eq: xi}
\xi=\max \{1,1/2\ln (y_{\rm max})[1.6+\ln (y_{\rm max})]\},
\end{equation}
where $y_{\rm max}\approx 3(\rho/10^{-9}~{\rm g~cm^{-3}})^{-1/2}(T/100 {\rm eV})^{9/4}$. Taking into account the Comptonization, the temperature will be modified as,
\begin{equation}
\label{eq: T_Comp}
T_{\rm Comp}=T_{\rm BB} \left\{
\begin{aligned}
&\eta^2/\xi (T_{\rm Comp})^2, &\eta>1 \\
&1, &\eta<1
\end{aligned}
\right.
\end{equation}

In addition, the pair producing process would prevent the radiation exceeding the maximum temperature, e.g., $\sim 100\ \rm \rm keV$. Therefore, the temperature of the cocoon breakout can be defined as,
\begin{equation}
\label{Eq: T_c}
T_{\rm c}=\min \{ T_{\rm Comp}, 100\ \rm keV\}.
\end{equation}

\section{Light Curves, Spectra and Observability} \label{results}
Now, we can calculate the emission from a GRB with two-component jet in AGN disk by considering the cavity produced by different progenitor wind model.

The parameters and values we adopted for the AGN disk and two-component jet are described in Table \ref{tab:parameter}: the SMBH mass $M_{\rm SMBH}$, the AGN disk height $H$, the midplane density of AGN disk $\rho_{\rm d}$, the sound speed of disk $c_{\rm s}$, the location of GRB in AGN disk $R_{\rm GRB}$, the cavity size $r_{\rm cav}$, the duration of GRB $t_{\rm j}$, the initial Lorentz factor $\Gamma_0$, the initial half-opening angle $\theta_{\rm j0}$, the isotropic kinetic energy $E_{\rm 0,iso}$, the fraction of the shock energy into electrons $\epsilon_{\rm e}$ and magnetic field $\epsilon_{\rm B}$, and the electron energy distribution index $p$.
Besides, in order to eliminate the bias of parameter selection, we adopt three different groups of $E_{\rm 0,iso}$ for the narrow ($E_{\rm n,iso}$) and wide ($E_{\rm w,iso}$) component, $E_{\rm w,iso}=0.1E_{\rm n,iso}$, $E_{\rm w,iso}=E_{\rm n,iso}$ and $E_{\rm w,iso}=10E_{\rm n,iso}$, where $E_{\rm n,iso}=2\times 10^{52} {\rm erg~s^{-1}}$.

\begin{table}
\centering \caption{Parameters of AGN disk and GRB two-component jet}
\begin{tabular}{ccc}
\hline
\multicolumn{3}{c}{AGN disk}                                                                                 \\ \hline
\multicolumn{1}{c|}{$M_{\rm   SMBH}~(M_\odot)$}    & \multicolumn{2}{c}{$10^8$}                              \\
\multicolumn{1}{c|}{$H ({\rm cm})$}                        & \multicolumn{2}{c}{$10^{14}$}                           \\
\multicolumn{1}{c|}{$\rho_{\rm   d}~({\rm g~cm^{-3}})$} & \multicolumn{2}{c}{$10^{-10}$}                          \\
\multicolumn{1}{c|}{$c_{\rm s}~   ({\rm cm~s^{-1})}$}    & \multicolumn{2}{c}{$5\times 10^{6}$}                    \\ \hline
\multicolumn{3}{c}{GRB and progenitor}                                                                       \\ \hline
\multicolumn{1}{l}{}                               & SGRB              & LGRB                                \\ \hline
\multicolumn{1}{c|}{$R_{\rm   GRB}~(R_{\rm s})$}   & 500               & 1000                                \\
\multicolumn{1}{c|}{$r_{\rm   cav}~(H)$}           & 1                 & 0.05                                \\
\multicolumn{1}{c|}{$t_{\rm   j}~({\rm s})$}             & 2                 & 30                                  \\
\multicolumn{1}{c|}{$r_{\rm   ph}~(H)$}            & 0.27              & 1.1                                 \\ \hline
\multicolumn{3}{c}{Two-component jet}                                                                        \\ \hline
\multicolumn{1}{l}{}                               & narrow            & wide                                \\ \hline
\multicolumn{1}{c|}{$\Gamma_0$}                    & $300$             & $30$                                \\
\multicolumn{1}{c|}{$\theta_{\rm   j0}$}           & $5^\circ$         & $10^\circ$                          \\
\multicolumn{1}{c|}{$E_{0,iso}~({\rm erg})$}             & $2\times 10^{52}$ & $2\times [10^{51},10^{52},10^{53}]$ \\
\multicolumn{1}{c|}{$\epsilon_{\rm   e}$}          & $0.3$             & $0.3$                               \\
\multicolumn{1}{c|}{$\epsilon_{\rm   B}$}          & $0.01$            & $0.01$                              \\
\multicolumn{1}{c|}{$p$}                           & $2.3$             & $2.3$                               \\ \hline
\end{tabular}
\label{tab:parameter}
\end{table}

\begin{figure*}
\centering
\includegraphics [angle=0,scale=0.3] {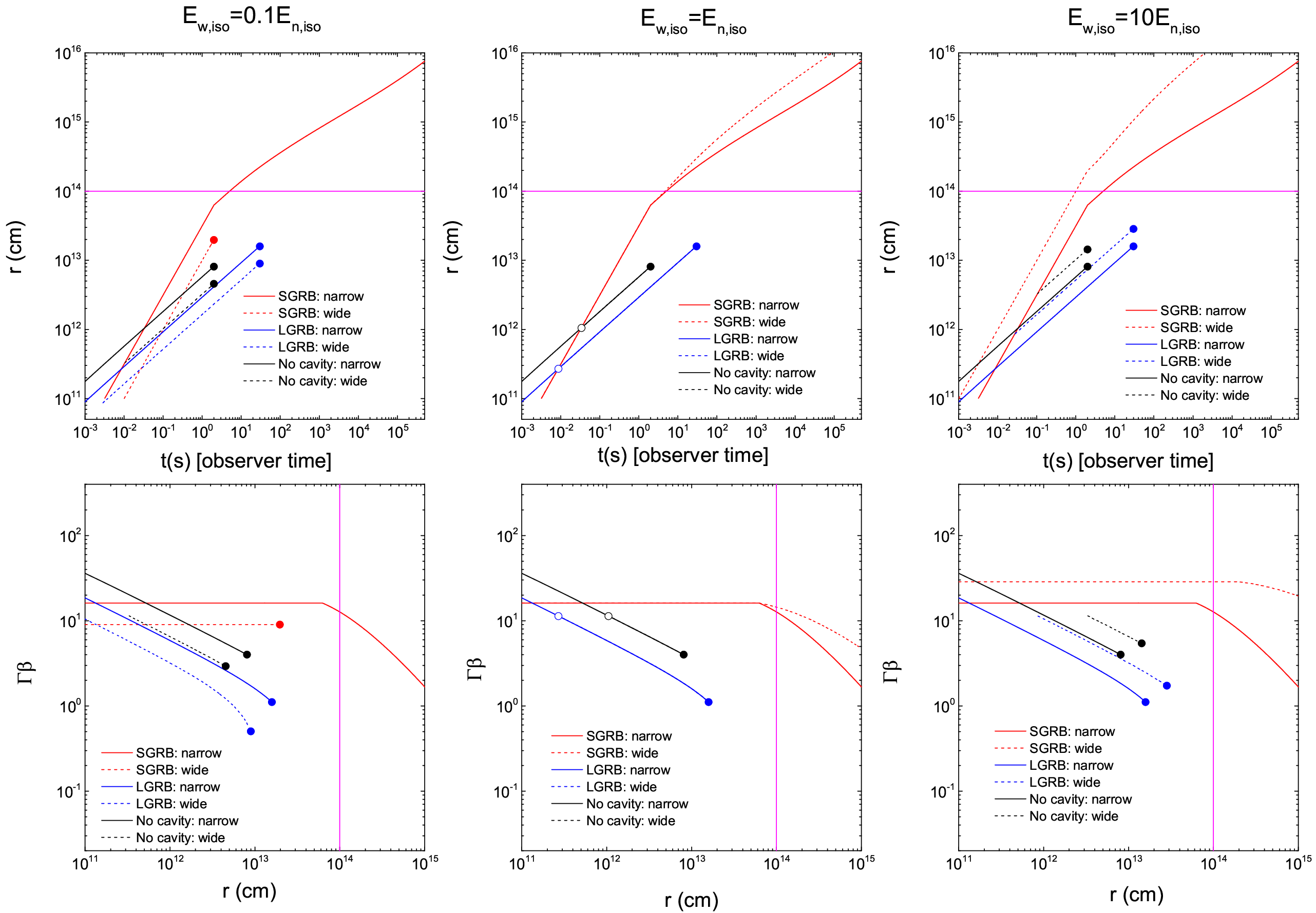}
\caption{Dynamics of the narrow (solid lines) and wide (dashed lines) jets in the AGN disk for three cases, the no-cavity case (black), the SGRB case (red) and the LGRB case (blue) with $E_{\rm w,iso}=0.1E_{\rm n,iso}$ (left panel), $E_{\rm w,iso}=E_{\rm n,iso}$ (middle panel) and $E_{\rm w,iso}=10E_{\rm n,iso}$ (right panel). Top panel: the time evolution of the jet radius $r$. The horizontal magenta line represents the AGN disk surface; Bottom panel: the evolution of $\Gamma\beta$ of the jet with $r$. The vertical magenta line represents the AGN disk surface.
The filled circles at the end of each line present the jet choked time or radius. We only present the dynamics beyond $r_{\rm N}$, where $r_{\rm N}=l^{3/2}/(\Delta^{1/2}\Gamma_0^2)$ is the radius at which the RS transitions from Newtonian to relativistic \citep{Zhang2018}. In the middle panels, to make it clearer, we use the open circles to mark the locations of $r_{\rm N}$ for the wide components (dashed lines) of LGRB (blue lines) and no-cavity (black lines) cases.}
\label{fig:dynamic_binary}
\end{figure*}

The dynamic evolution of the two-component jet (solid lines for narrow jet and dashed lines for wide one) are presented in Fig. \ref{fig:dynamic_binary} for SGRB and
LGRB. The evolution of radius $R$ and $\Gamma\beta$ are plotted in the left and right panels, respectivley. For comparison, we also show the results without cavity (black lines). We adopt $t_{\rm j} \sim 2$s (30s) for SGRB (LGRB) case.

For the SGRB case, as shown in section 2.2.1, the disk wind can penetrate the AGN disk in the vertical direction forming the cavity, and the cavity would be refilled by the AGN disk material. The GRB jet can occur during either the refilled stage or the penetrating stage. In our calculation, we assume that when the jet occurs, the AGN disk is penetrated by the disk wind. Therefore, we consider cavity size $r_{\rm cav}\sim H$ in the vertical direction.
For the LGRB case, $r_{\rm cav}\approx 0.05H$ is adopted. We find that the results for the LGRB case roughly resemble the case without cavity, for which both the narrow and wide jets will be choked according to Eq. (\ref{Eq: t_j_bo}). For SGRB case, except the wide component with $E_{\rm w,iso}=0.1E_{\rm n,iso}$, the jets would successfully break out from the photosphere of the wind, due to $t_{\rm j,bo}< 2\Gamma_{\rm j}^2t_{\rm j}$. We would thus expect a non-thermal emission for these scenarios.

\subsection{Light Curves}\label{lightcurve}
Based on the evolution of FS and RS, we can calculate the emergent emission from the GRB in AGN disk.

\begin{figure*}
\centering
\includegraphics [angle=0,scale=0.2] {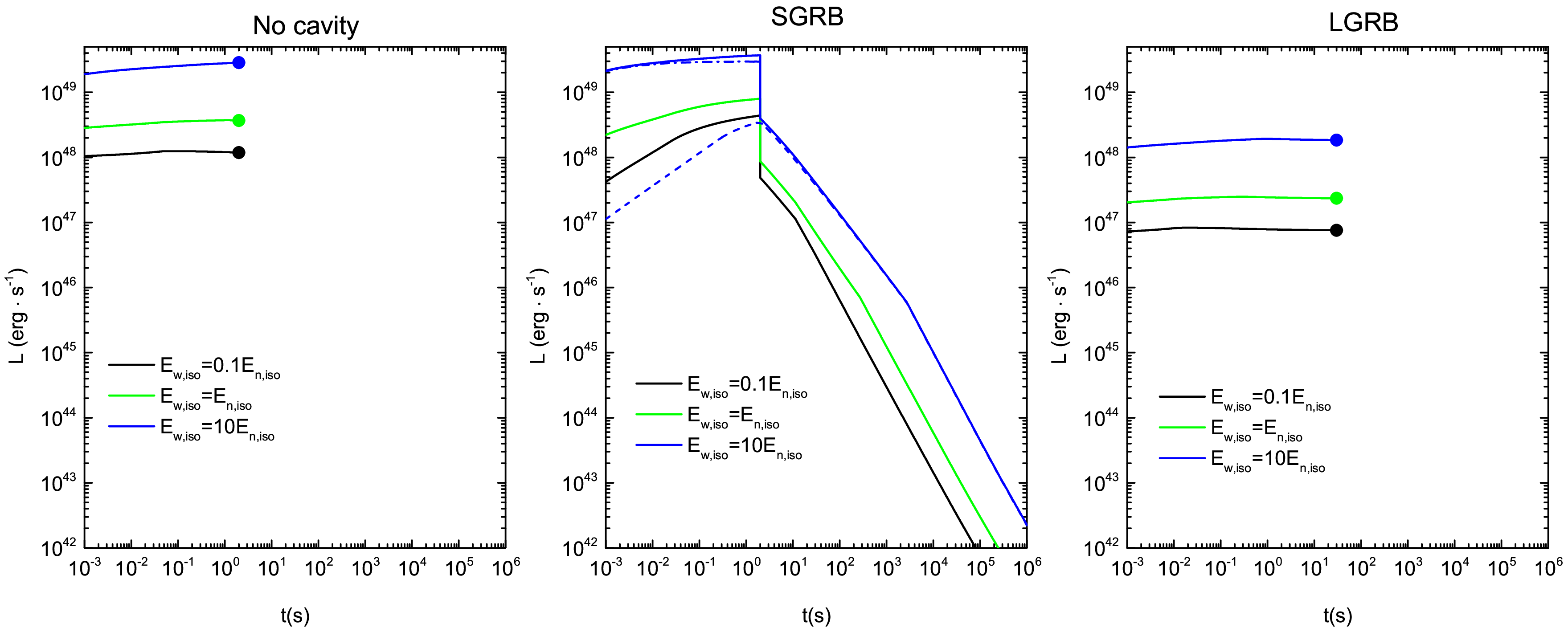}
\caption{
The panels show the total FS and RS luminosity for $E_{\rm w,iso}=0.1E_{\rm n,iso}$ (black line), $E_{\rm w,iso}=E_{\rm n,iso}$ (green line) and $E_{\rm w,iso}=10E_{\rm n,iso}$ (blue line). The filled circles at the end of lines represent the time that both the narrow and wide component are choked. Besides, we decompose the total radiation of $E_{\rm w,iso}=10E_{\rm n,iso}$ of SGRB into the contributions of FS (dashed blue line) and RS (dash-dot line) in the middle panel.
Three cases, the no-cavity (left), SGRB (middle) and LGRB (right) are considered.}

\label{fig:lc_case}
\end{figure*}

In Fig.\ref{fig:lc_case}, we present the total emission of FS and RS for no-cavity (left), SGRB (middle) and LGRB (right). And, due to the high density of the ambient medium, these emission is difficult to escape from the radiation region.
Besides, we also calculate the cocoon emission, which is not shown in the figure (the duration is too short, making them hard to distinguish in the figure).

It is found that, for both the FS/RS emission, the early emission is dominated by the RS emission and the late emission by the FS emission.The filled circles at the end of lines present the time that both the narrow and wide components are choked.
As discussed above, except the wide component with $E_{\rm w,iso}=0.1E_{\rm n,iso}$, the jets would successfully break out from the photosphere of the wind in the SGRB case.  We therefore expect to observe the non-thermal FS and RS emissions peaking at $5 \times 10^{48}~ \rm erg~s^{-1} $ in SGRB case.

The narrow (wide with $E_{\rm w,iso}=0.1, 1$ and $10 E_{\rm n,iso}$) cocoon emission for the cases of no-cavity and LGRB share similar characteristics, e.g.,
breakout time $\sim 18255$ s ($\sim$ 31908, 9708, 3372 s), duration $\sim 35$ s ($\sim$ 90, 10, 2 s) and luminosity $\sim 7\times 10^{45}~\rm  erg~s^{-1}$ ( $\sim 1\times 10^{45}, 5\times 10^{46}, ~{\rm and}~ 1\times 10^{48} ~\rm  erg~s^{-1}$) for no-cavity and breakout time $\sim 16328$ s ($\sim$ 30018, 6963, 2721 s), duration $\sim 36$ s ($\sim$ 100, 10, 2 s) and luminosity $\sim 7\times 10^{45}~\rm  erg~s^{-1}$ ( $\sim 1\times 10^{45}, 5\times 10^{46}, ~{\rm and}~ 1\times 10^{48} ~\rm  erg~s^{-1}$) for LGRB.
However, for the SGRB case, only the wide component with $E_{\rm w,iso}=0.1E_{\rm n,iso}$ can be choked, and its breakout time $\sim$ 100 s, duration $\sim$ 600 s and luminosity $\sim 8\times 10^{47} ~\rm  erg~s^{-1}$.
Therefore, compared with the LGRB case, the shock breakout time of cocoons in SGRB case is earlier, and the duration of radiation is slightly longer.

\subsection{Spectra and Observability}\label{sepctrum}

\begin{figure*}
\centering
\includegraphics [angle=0,scale=0.2] {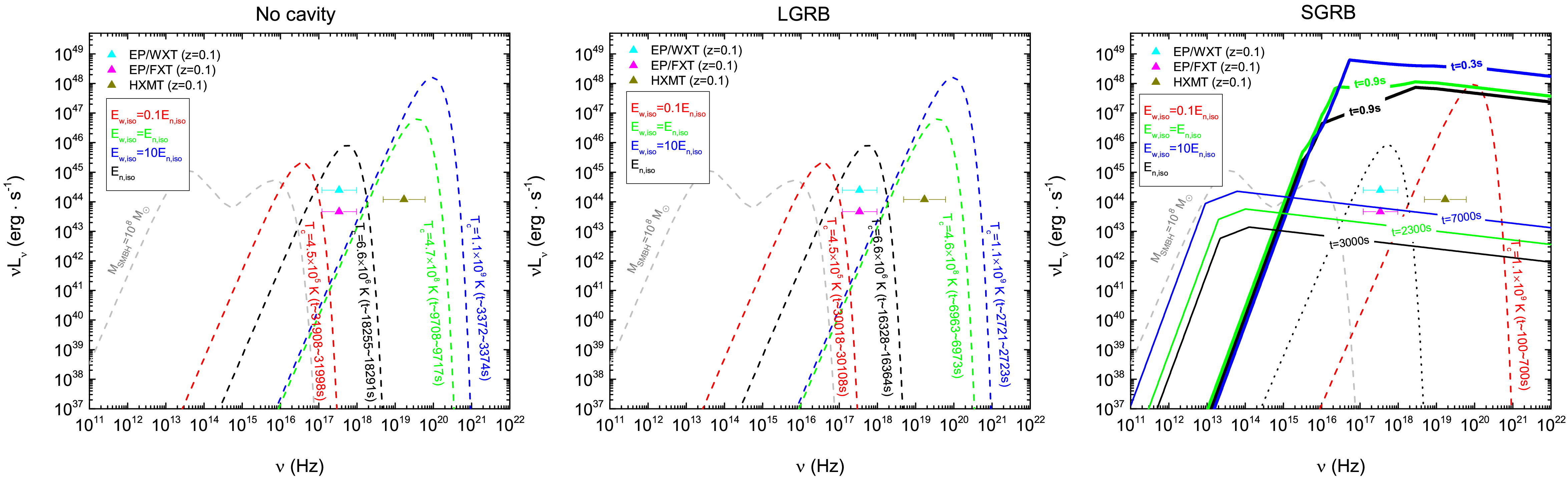}
\caption{The spectra of the wide cocoon emission for no-cavity (left panel) and LGRB (middle panel) cases with $E_{\rm w,iso}=0.1E_{\rm n,iso}$ (red dashed lines), $E_{\rm w,iso}=E_{\rm n,iso}$ (green dashed lines) and $E_{\rm w,iso}=10E_{\rm n,iso}$ (blue dashed lines). The spectra for the narrow cocoon component are plotted with black dashed lines in these two panels. For comparison, the AGN disk background are plotted with gray dashed lines. In the right panel (SGRB case), the afterglow emission from the narrow component and the cocoon emission from the wide component for $E_{\rm w,iso}=0.1E_{\rm n,iso}$ case are plotted with black solid lines and red dashed line, respectively. The afterglow spectrum (combining the narrow and wide components) for $E_{\rm w,iso}=1E_{\rm n,iso}$ (green solid lines) and $E_{\rm w,iso}=~10E_{\rm n,iso}$ (blue solid lines) are also shown. The result of \citet{Zhuetal.(2021)} is shown with black dotted line. The timestamps ($t=0.9$s, $0.9$s and $0.3$s) on the thick solid lines represent the time of the successful breakout of the SGRB jets for these three cases. We also plot the afterglow emissions at late time (thin solid lines) for comparison. The different color triangles represent the sensitivity of EP/WXT, EP/FXT and HXMT, respectively.
}
\label{fig:spec}
\end{figure*}


We further analyze the spectral characteristics and compare with that of the AGN disk. The results are shown in Fig. \ref{fig:spec}. We have computed the spectra of the AGN disk $M_{\rm SMBH}=10^8 M_\odot$.

We also calculate the possibility of detection in X-ray, soft gamma-ray, and optical bands by comparing with the thresholds of EP/WXT \citep{Yuan(2017)} EP/FXT\citep{Yuanetal.(2015), Yuanetal.(2022)} and HXMT \citep{Li(2007)} in Fig. \ref{fig:spec}, respectively.

We found that, the cocoon spectra for the LGRB is similar to those of the case without cavity. The temperatures of the wide cocoon emissions are labeled in Fig. \ref{fig:spec}, i.e.,  $\sim 4.5\times 10^5$ ($4.6\times 10^8$, $1.1\times 10^9$) K for $E_{\rm w,iso}=0.1$ (red dashed lines), 1 (green dashed lines) and  $10 E_{\rm n,iso}$ (blue dashed lines). For narrow cocoon of LGRB, the temperature is $\sim 6.6\times 10^6$ K (black dashed lines).
We also present the EP/WXT (cyan triangle), EP/FXT (magenta triangle) and HXMT (dark yellow triangle) sensitivity. The cocoon emission can be detected by EP and HXMT at redshift $z\lesssim 0.1$.

The radiation characteristics of SGRB are unique (see the right panel of Fig. \ref{fig:spec}). For comparison, we also show the narrow cocoon emission from the case without cavity (black dotted line), which is just the result of \citet{Zhuetal.(2021)}. The temperature of cocoon breakout emission in SGRB case is much higher than the no-cavity case, which can be monitored by HXMT. Besides, except the wide component with $E_{\rm w,iso}=0.1E_{\rm n,iso}$, the jets would successfully break out from the photosphere of the circum-binary disk wind and contribute a non-thermal emission. In the right panel, the timestamps ($t=0.9$s, $0.9$s and $0.3$s) on the thick solid lines represent the time of the successful breakout of the SGRB jets for $E_{\rm w,iso}=0.1E_{\rm n,iso}$ (black), $E_{\rm w,iso}=E_{\rm n,iso}$ (green) and $E_{\rm w,iso}=10E_{\rm n,iso}$ (blue), respectively. For comparison, we also present the afterglow emissions at late time (thin solid lines).
The jet non-thermal emission all can be detected by EP/FXT and HXMT at $z\lesssim 0.1$.

\section{Conclusions and Discussions}\label{Conclusions}
In this work, we investigate the emission of a GRB with two-component jet in the AGN disk by considering the effects of progenitor wind on density distribution. In addition, we adopt three different scenarios for isotropic kinetic energy, $E_{\rm w,iso}=0.1$, $1$ and $10E_{\rm n,iso}$, to eliminate the bias of parameter selection.

Our conclusions are summarized as follows:

1. Both progenitors (SGRBs and LGRBs) will develop a cavity in the AGN disk. The strong wind from the SGRB progenitors can penetrate the disk surface, while the LGRB progenitor can only produce a relatively small cavity.

2. All the narrow and wide jets of the LGRBs will be choked in the disk, and the dynamical behaviors are similar to the case without cavity. For SGRBs, except the wide component with $E_{\rm w,iso}=0.1E_{\rm n,iso}$, the jets would successfully break out from the photosphere of the BCOs wind.

3. The emissions from the LGRBs and the no-cavity cases are similar. The SGRBs can produce broad cocoon-breakout emission (the wide component with $E_{\rm w,iso}=0.1E_{\rm n,iso}$), and non-thermal emission except the wide component with $E_{\rm w,iso}=0.1E_{\rm n,iso}$.

4. For all the cases, the cocoon and non-thermal emission is much brighter. For the LGRBs and the case without cavity, the cocoon breakout emission can be detected by EP and HXMT. For the SGRBs in disk, the jet afterglow (non-thermal) emission and cocoon breakout emission from narrow and wide jets can be monitored by EP and HXMT, respectively.

Therefore, the joint observations by EP and HXMT might be helpful to distinguish the type of GRBs in AGN disk and the jet components.

We have assumed that the progenitor winds are isotropic, and the jets are launched from the middle plane and perpendicular to the AGN disk. However, the wind could be angle dependent, which may be weak in polar direction for the SGRB case. The location of the jet launching and its orientation could be random in disk. The two-component jets launched near disk surface would produce a cocoon breakout emission and a non-thermal afterglow emission similar to the SGRB case discussed in this paper. The inclined jets or the precessing jet \citep{Lei2007, Liu2010, Gao2023} would be difficult to break out of the disk surface, even in the SGRB case. The numerical simulations would be helpful for detailed investigations \citep{Kathirgamarajuetal.(2023),Martin2024}. \citet{Mimica2015} conducted numerical simulation of the two-component jet in tidal-disruption events. The two-component jet launching in a dense medium (e.g., in AGN disk) should be explored in future numerical simulations.

We adopted duration $t_{\rm j}=2$s for SGRBs. Traditionally, $t_{\rm j} \sim 2$s is taken as the separation line for long and short GRBs. However, there are some merger-type GRBs with duration $t_{\rm j} >2$s \citep{Gao2022}. Therefore, $t_{\rm j}=2$ could be a reasonable value for SGRBs. It should be noted that the GRBs may have late time central engine activities \citep{Wu2013,Gao2016,Chen2017,Ma2018,Zhao2021,Huangetal.(2024)}, which would enhance the emissions \citep{Huang2004}.

In addition to GRBs, there are other transients, e.g., SNe \citep{Lietal.(2023)}, FRB \citep{Zhaoetal.(2024)}, tidal disruption events \citep{Prasad2024} and X-ray transients \citep{ChenDai(2023), Tagawaetal.(2023), Tagawaetal.(2024)}.
Identification of GRBs in AGN disk is still challenging, especially considering other events with similar characteristics like tidal disruption events.
However, strong gravitational wave signals are expected for these catastrophic events. In the future, we anticipate that ground and space gravitational wave detectors with higher sensitivity, such as the Einstein Telescope \citep{Punturoetal.(2010a), Punturoetal.(2010b)}, Laser Interferometer Space Antenna \citep{Amaro-Seoaneetal.(2017)}, and TianQin \citep{Amaro-Seoaneetal.(2017)}, will detect more gravitational wave signals from AGN disks. Furthermore, neutrinos bursts could be detected by IceCube following the observation of gravitational wave \citep{Zhu2024}. Therefore, multi-messenger follow-up observations are crucial for verifying the model proposed in this paper.

\begin{acknowledgements}
We are very grateful to Yaping Li, Hui Li, Fulin Li, Bing Zhang, He Gao and Yunwei Yu for their helpful discussions. This work is supported by the National Key R\&D Program of China (Nos. 2020YFC2201400, 2023YFC2205901), and the National Natural Science Foundation of China under grant 12473012. W.H.Lei. acknowledges support by the science research grants from the China Manned Space Project with NO.CMS-CSST-2021-B11.
\end{acknowledgements}

\bibliography{export-bibtex}{}

\end{document}